\documentclass[hidelinks,onefignum,onetabnum]{siamart220329}

\usepackage{amsmath}
\usepackage{algorithmic}

\usepackage{subcaption}

\usepackage{pgfplots}
\usepackage{pgfplotstable}
\usetikzlibrary{decorations.markings}
\usetikzlibrary{patterns}
\usetikzlibrary{graphs}
\usetikzlibrary {matrix}
\pgfplotsset{compat = 1.13,
        my ybar legend/.style={
            legend image code/.code={
                \draw [##1] (0cm,-0.6ex) rectangle +(2em,1.5ex);
            },
        },
}

\author{Thijs Steel\thanks{KU Leuven, Belgium
}
\and Julien Langou\thanks{University of Colorado Denver, USA and Centre Inria de Lyon, France
}}

\title{Communication efficient application of sequences of planar rotations to a matrix}

\date{April 2024}

\begin{document}

\maketitle

\begin{abstract}
We present an efficient algorithm for the application of sequences of planar
rotations to a matrix. Applying such sequences efficiently is important in many
numerical linear algebra algorithms for eigenvalues.
Our algorithm is novel in three main ways.
First, we introduce a new kernel that is optimized for register reuse in a novel way.
Second, we introduce a blocking and packing scheme that improves the cache efficiency of the algorithm.
Finally, we thoroughly analyze the memory operations of the algorithm which leads to important
theoretical insights and makes it easier to select good parameters.
Numerical experiments show that our algorithm outperforms the state-of-the-art and achieves a flop
rate close to the theoretical peak on modern hardware.
\end{abstract}


% REQUIRED
% \begin{MSCcodes}
%    65F15, 65F25, 65Y20
% \end{MSCcodes}

\definecolor{ferngreen}{HTML}{56641a}
\definecolor{perfumepurple}{HTML}{c0affb}
\definecolor{apricotorange}{HTML}{e6a176}
\definecolor{orientblue}{HTML}{00678a}
\definecolor{winered}{HTML}{984464}
\definecolor{downygreen}{HTML}{5eccab}

\section{Introduction}\label{section:introduction}

The problem we will focus on in this paper is the application of a sequence of
planar rotations to a matrix. A basic algorithm to apply such a sequence can be
simple: we just loop over the rotations and apply each one to the matrix.
Pseudocode for this algorithm is given in Algorithm~\ref{alg:rot_sequence} and
our C implementation is not much more complicated.

Applying a sequence of orthogonal transformations efficiently is an important
building tool in numerical linear algebra. To achieve high performance, many
factorizations limit their initial calculations to a smaller submatrix of the
original matrix. Updating the rest of the matrix (which often involves the bulk
of the floating-point operations) can then be done efficiently with an
optimized routine. The two most common orthogonal transformations are Givens
rotations and Householder reflectors. Givens rotations are applied to two
vectors (usually two columns or rows of a matrix) and are defined by a cosine
and a sine. Householder reflectors can involve an arbitrary number of vectors.
To apply a sequence of reflectors efficiently, the most popular method relies
on the WY representation~\cite{schreiber1989storage}. Through the use of this
factorization, routines that apply a sequence of reflectors to a matrix can be
optimized to achieve close to the theoretical peak flop rate on modern
hardware. Because of this, reflectors have become the preferred choice for
orthogonal transformations. However, some algorithms require the use of
rotations. In case the matrix has a special structure (for example, if it is
upper triangular), applying a reflector to it can destroy that structure. If
rotations are used instead, the structure can more easily be preserved. Some
examples are the implicit QR algorithm~\cite{francis1961qr} and the Jacobi
method for the singular value decomposition~\cite{jacobi1846leichtes}.

The basic algorithm presented in Algorithm~\ref{alg:rot_sequence} is not very
efficient on modern hardware. The main reason for this is inefficient memory
access because the algorithm is not cache-friendly. If we refer to the rotation
whose values are stored in position $(i,j)$ of the matrices $C$ and $S$ as
rotation $(i,j)$, then we can say that column $i$ of the matrix the rotations
are applied to is involved in rotations $(i-1, j)$ and $(i,j)$ for all $j$. But
in between applying $(i,j)$ and $(i,j+1)$, we need to load and store the entire
matrix, so it is unlikely that the values of column $i$ are still in the cache
when we need them again. K{\aa}gstr{\"o}m et al.~\cite{kaagstrom2008blocked}
and later Van Zee et al.~\cite{van2014restructuring} have improved upon this
basic algorithm in two ways, fused rotations and a wavefront pattern. 

\subsection{Wavefront algorithm}\label{subsec:wavefront} 
\begin{figure}
    \centering
	\subcaptionbox{Standard pattern}[.49\textwidth]{%
    \centering
    \begin{tikzpicture}
        \draw (0,0) rectangle (1,3);
        \foreach \x in {0,...,3}
        \draw[decoration={markings,mark=at position 0.5 with {\arrow{>}}}, postaction={decorate}] (\x/4 + 0.5/4,2.9) -- (\x/4 + 0.5/4,0.1);

        % \foreach \x in {0,...,2}
        % \draw[decoration={markings,mark=at position 0.5 with {\arrow{>}}}, postaction={decorate}] (\x/4 + 0.125,0.1) -- (\x/4 + 0.375,2.9);

    \end{tikzpicture}
    }
	\subcaptionbox{Wavefront pattern}[.49\textwidth]{%
    \centering
    \begin{tikzpicture}
        \draw (0,0) rectangle (1,3);

        \draw[decoration={markings,mark=at position 0.5 with {\arrow{>}}}, postaction={decorate}] (0.125,2.9-0.25) -- (0.125 + 0.25,2.9);

        \draw[decoration={markings,mark=at position 0.5 with {\arrow{>}}}, postaction={decorate}] (0.125,2.9-0.5) -- (0.125 + 0.5,2.9);

        \foreach \x in {0,...,8}
        \draw[decoration={markings,mark=at position 0.5 with {\arrow{>}}}, postaction={decorate}] (0.125,2.9-0.75-\x/4) -- (0.125 + 0.75,2.9-\x/4);

        \draw[decoration={markings,mark=at position 0.5 with {\arrow{>}}}, postaction={decorate}] (0.125+0.25,0.125) -- (0.875,0.125+0.5);

        \draw[decoration={markings,mark=at position 0.5 with {\arrow{>}}}, postaction={decorate}] (0.125+0.5,0.125) -- (0.875,0.125+0.25);

    \end{tikzpicture}
    }
    \caption{The matrix $C$ containing the cosines of the rotations and arrows indicating the order in which the rotations are applied. On the left, the standard pattern which applies full sequences of rotations. On the right, the wavefront pattern which applies the rotations in ``waves''.}
    \label{fig: wavefront}
\end{figure}
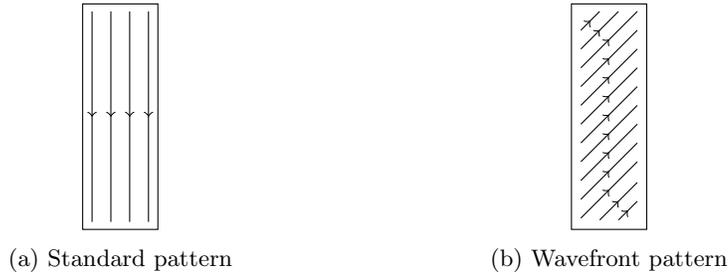

In the standard
algorithm, we need to load and store the entire matrix between applying
rotation $(i,j)$ and $(i,j+1)$. In an ideal world, we would just change the
order of the loops so that rotation $(i,j+1)$ is applied immediately after
rotation $(i,j)$. Sadly, this is not allowed. Before we can apply rotation
$(i,j+1)$, we have to apply rotation $(i+1,j)$. This leads to a wavefront
pattern, where we apply rotation $(i,j)$, then $(i-1,j+1)$, $(i-2,j+2)$,
$\dots$. Figure~\ref{fig: wavefront} shows this pattern visually; each of the
diagonal sequences is called a wave. The advantage of this pattern is that
instead of accessing $n-1$ columns before a column is used again, this pattern
accesses $k$ columns. Typically, $k$ is much smaller than $n$, so it is much
more likely that the column can remain in the cache.

\subsection{I/O complexity}

If we assume a two-memory machine with a small cache of size $S$ and a large
memory, we want to quantify the number of memory movement (read and write)
between the memory and the cache. This is called the I/O complexity problem. 

Recent work on I/O lower bounds have led to techniques and tools to
mechanically compute the I/O lower bounds of parametrized
programs~\cite{olivry2020automated}. Using the IOLB tool on
Algorithm~\ref{alg:rot_sequence} leads to a I/O lower bound of $\frac{ m n
k}{\sqrt{S}}.$ Since the total number of operations of our code is $6mnk$, this
means that the operational intensity is at most $6\sqrt{S}$.

A quick analysis of the wavefront algorithm explained in
Section~\ref{subsec:wavefront} goes as follows.  
We need to apply $mnk$ Givens rotations.        % $m(n-1)k$ for real We find parameters
$m_b$ and $k_b$ such that $  m_b k_b  \leq S $  % $ k_b  m_b + 2 \leq S $ for
real such that a block of size $m_b$-by-$k_b$ of $A$ fits in cache.  At each
step of the wavefront algorithm, we need to read one column of size $m_b$, and
write back ne column of size $m_b$, and load $2 k_b$ cosines and sinces that
represent $k_b$ rotations.  In this step, we will be able to do $  m_b k_b  $
rotations.  Forgetting start and end clean up code, we need to do (roughly) $
\frac{ m n k}{m_b k_b}$ steps.  And so the I/O of the algorithm is $ \frac{ mnk
}{ m_b  k_b }  ( 2m_b + 2k_b ).$ For $k_b = \sqrt{S}$ and  $mb = \sqrt{S}$ we
get $\frac{4 m n k}{\sqrt{S}}.$ Since the total number of operations of our
code is $6mnk$, this means that the operational intensity for the wavefront
algorithm is $\frac{3}{2}\sqrt{S}$.

Two conclusions. First, we see that the ratio between an I/O lower bound for
our problem and the wavefront algorithm is a factor 4. Second, we see that the
operational intensity of our kernel is pretty good. (For comparison,  the
operational intensity of GEMM is $\sqrt{S}$.) Therefore there is potential to
be able to hide I/O's and reach GEMM-like performance.

We note that an analysis of the number of cache misses of the wavefront
algorithm, a related problem, is done in~\cite[\S4.2]{van2014restructuring}.

\subsection{Fused rotations} Despite what the name suggests, fused rotations
are not directly accumulated in any way. Instead, a fused rotation is a
technique to increase register reuse. The advantage becomes clear when we study
what happens when we apply two rotations to three vectors $x$, $y$, and $z$:
\begin{enumerate}
    \item Load rotation 1 into registers
    \item for $i = 0, 1, \dots, n$
    \item \hspace{2em}load $x[i]$ and $y[i]$ into registers
    \item \hspace{2em}apply rotation 1 to $x[i]$ and $y[i]$
    \item \hspace{2em}store $x[i]$ and $y[i]$
    \item Load rotation 2 into registers
    \item for $i = 0, 1, \dots, n$
    \item \hspace{2em}load $y[i]$ and $z[i]$ into registers
    \item \hspace{2em}apply rotation 2 to $y[i]$ and $z[i]$
    \item \hspace{2em}store $y[i]$ and $z[i]$
\end{enumerate}
A fused rotation would change this to:\begin{enumerate}
    \item Load both rotations into registers
    \item for $i = 0, 1, \dots, n$
    \item \hspace{2em}load $x[i]$, $y[i]$, and $z[i]$ into registers
    \item \hspace{2em}apply rotation 1 to $x[i]$ and $y[i]$
    \item \hspace{2em}apply rotation 2 to $y[i]$ and $z[i]$
    \item \hspace{2em}store $x[i]$, $y[i]$, and $z[i]$
\end{enumerate}
This applies the exact same arithmetic operations, but only loads and stores
$y[i]$ once instead of twice. This is important because memory operations are
much more expensive than arithmetic operations on modern hardware. We can use
fused rotations to fuse rotations $(i,j)$ and $(i+1,j)$ or rotations $(i,j)$
and $(i-1,j+1)$. We refer to these as $2 \times 1$ fused rotations and $1
\times 2$ rotations, respectively. We can fuse even more rotations so long as
we have enough registers to store all the values. On a typical CPU, it is
optimal to use $2 \times 2$ fused rotations.

\begin{algorithm}[]
    \caption[rot]{rot($x$, $y$, $c$, $s$)\\Apply a single rotation to two vectors.}
    \label{alg:rot}
    \begin{algorithmic}[1]
        \REQUIRE Vectors $x$ and $y$ of length $m$, scalars $c$ and $s$
        \FOR{$(i = 0; i < m; i++)$}
        \STATE{$t = c \cdot x[i] + s \cdot y[i]$}
        \STATE{$y[i] = -s \cdot x[i] + c \cdot y[i]$}
        \STATE{$x[i] = t$}
        \ENDFOR
    \end{algorithmic}
\end{algorithm}

\begin{algorithm}[]
    \caption[rot\_sequence]{rot\_sequence($A$, $C$, $S$)\\
        Apply a sequence of $(n-1) k$ rotations, stored in $C$ and $S$, to a matrix $A$ from the right.}
    \label{alg:rot_sequence}
    \begin{algorithmic}[1]
        \REQUIRE Matrix $A$ of size $m \times n$, matrices $C$ and $S$ of size $(n-1) \times k$
        \FOR{$(p = 0; p < k; p++)$}
        \FOR{$(j = 0; j + 1< n; j++)$}
        \STATE{rot($A(:,j)$, $A(:,j+1)$, $C(j,p)$, $S(j,p)$)}
        \ENDFOR
        \ENDFOR
    \end{algorithmic}
\end{algorithm}

\begin{algorithm}[]
    \caption[rot\_sequence\_wavefront]{rot\_sequence\_wavefront($A$, $C$, $S$)\\
        Apply a sequence of $(n-1) k$ rotations, stored in $C$ and $S$, to a matrix $A$ from the right. Note: this algorithm assumes $k \leq n-1$.}
    \label{alg:rot_sequence_wavefront}
    \begin{algorithmic}[1]
        \REQUIRE Matrix $A$ of size $m \times n$, matrices $C$ and $S$ of size $(n-1) \times k$
        \FOR{$(p = 0; p < k-1; p++)$ \COMMENT{Startup phase}}
        \FOR{$(l = 0, j = p; l < j+1; ++l,--j)$}
        \STATE{rot($A(:,j)$, $A(:,j+1)$, $C(j,p)$, $S(j,p)$)}
        \ENDFOR
        \ENDFOR
        \FOR{$(p = k-1;p<n-1;++p)$\COMMENT{Pipeline phase}}
        \FOR{$(l=0;j = p;l < k; ++l,--j)$}
        \STATE{rot($A(:,j)$, $A(:,j+1)$, $C(j,p)$, $S(j,p)$)}
        \ENDFOR
        \ENDFOR
        \FOR{$(p = n-k;p<n-1;++p)$\COMMENT{Shutdown phase}}
        \FOR{$(l=1;j = n-2;l < k; ++l,--j)$}
        \STATE{rot($A(:,j)$, $A(:,j+1)$, $C(j,p)$, $S(j,p)$)}
        \ENDFOR
        \ENDFOR
    \end{algorithmic}
\end{algorithm}

\subsection{Structure of this paper} By combining fused rotations and the
wavefront pattern, Van Zee et al.~\cite{van2014restructuring} were able to
achieve close to the theoretical peak flop rate on their machine. However,
since they published their paper, computer architectures have advanced and
memory operations have become more expensive relative to floating point
operations. In this paper, we will introduce several techniques to further
reduce the number of memory operations so that the algorithm can achieve
optimal flop rates even on modern machines. In Section~\ref{section:blocking},
we will expand upon the wavefront pattern and introduce a blocked version of
the algorithm which can efficiently utilize the caches. In
Section~\ref{section:kernel}, we will introduce a new technique to improve
register reuse. We will construct an efficient kernel that applies $n_r \times
k_r$ rotations to $m_r$ rows. Instead of keeping the rotations in registers and
loading new values of the columns in each iteration, we will keep some of the
values of the columns in registers and load new rotations in each iteration. We
will show that this significantly reduces the amount of memory operations.

\section{Blocking}\label{section:blocking}

Blocking is a commonly used technique to improve cache utilization. Instead of
working on a large matrix that does not fit into cache, we split the algorithm
into smaller pieces. Those smaller pieces will involve smaller matrices that do
fit into the cache, improving cache utilization.

The original algorithm applies $k$ sequences of $n-1$ rotations to $m$ rows of
a matrix. However, we cannot just split the algorithm into rectangular blocks
that apply $k_b$ sequences of $n_b$ rotations to $m_b$ rows. This is because of
the same application order restrictions that led to the wavefront pattern.
Inspired by the wavefront algorithm, we will instead split the algorithm into
blocks that apply $n_b$ waves of $k_b$ rotations to $m_b$ rows.
Figure~\ref{fig:fullalg} shows how the matrices are split into blocks and
Algorithm~\ref{alg:rot_sequence_block} shows an example implementation of such
a block. Even now that the rotations are split into blocks, we still need to
respect the order in which the rotations are applied. We need to apply the
block starting at $(i_b, j_b, p_b)$ before we can apply the block at $(i_b,
j_b+n_b, p_b)$ or the block at $(i_b, j_b-k_b, p_b+k_b)$. This mostly has
consequences when we will parallelize the code. It is also noteworthy that
while most of the blocks are parallelograms, some blocks are triangular. These
correspond to the startup and shutdown phases of the wavefront algorithm.

In Section~\ref{section:looporder}, we will discuss the order in which we
should treat the blocks and select the block sizes. But first, we will discuss
how to implement the block.

\begin{algorithm}[]
    \caption[rot\_sequence\_block]{rot\_sequence\_block($A$, $C$, $S$)\\
        Apply $n-k$ waves of $k$ rotations, stored in $C$ and $S$, to a matrix $A$ from the right. This is one block of the blocked algorithm. Note how this algorithm does not have a startup or shutdown phase.}
    \label{alg:rot_sequence_block}
    \begin{algorithmic}[1]
        \REQUIRE Matrix $A$ of size $m \times (n+k)$, matrices $C$ and $S$ of size $(n + k -1) \times k$
        \FOR{$(p = 0; p < k; p++)$}
        \FOR{$(j = k-1-p; j < n-p; j++)$}
        \STATE{rot($A(:,j)$, $A(:,j+1)$, $C(j,p)$, $S(j,p)$)}
        \ENDFOR
        \ENDFOR
    \end{algorithmic}
\end{algorithm}

\section{A kernel for register reuse}\label{section:kernel}

In the previous section, we discussed how to split the algorithm into blocks to
improve cache utilization. In this section, we will explain how to implement
those small blocks efficiently. Just like in the previous section, our main
focus will be reducing the cost of the memory operations, but instead of
improving cache utilization, we will focus on improving register reuse.

Algorithm~\ref{alg:rot_sequence_block} implements one of the small blocks. If
we count the number of memory operations needed, we see that we need $2
m_b(n_b-k_b) k_b$ loads and stores for the values in $A$, and $2(n_b-k_b)k_b$
loads for the values in $C$ and $S$ or
\begin{equation}
    4 m_b(n_b-k_b) k_b + 2(n_b-k_b)k_b \text{  memory operations.}
\end{equation}
In this calculation, we assume the values of $C$ and $S$ are loaded into
registers once and then reused for an entire rotation.

Using $2\times 2$ fused rotations, we can reduce the number of memory
operations to 
\begin{equation}
    2 m_b(n_b-k_b) k_b + 2(n_b-k_b)k_b \text{  memory operations.}
\end{equation}
The more rotations we can fuse, the more we can reduce the number of memory
operations. If we use $n_r \times k_r$ fused rotations, we can reduce the
number of memory operations to
\begin{equation}\label{eq:memops_fused}
    (\frac{2}{n_r} + \frac{2}{k_r} + \frac{2}{m_b}) m_b(n_b-k_b)k_b \text{  memory operations.}
\end{equation}
Unfortunately, these fused rotations also require us to fit $2 n_r k_r + k_r +
n_r$ values in registers (and we also need at least one register as temporary
storage to perform the rotation). On the CPUs we are targeting, we have 16
vector registers, so the largest fused rotation we can use is $2 \times 2$.

To reduce the number of memory operations further, we need a different
approach. We have mentioned before that we reuse the values of $C$ and $S$ by
keeping them in registers. In a $2 \times 2$ kernel, we can reuse $8$
registers, while we only need to load and store $6$ registers worth of data in
$A$. On modern CPUs, we can use vector registers, which can store multiple
values. On our machine, we can store 4 double-precision values in one vector
register. An implementation detail when using vector registers and instructions
is that we need to broadcast the values of $C$ and $S$ to registers, i.e. the
registers that contain cosines and sines do not contain 4 different cosines and
sines, but the same cosine and sine repeated 4 times. If we then look once
again at the memory reuse, we see that $8$ values of $C$ and $S$ can be reused,
while $24$ values of $A$ need to be loaded and stored. This ratio becomes even
worse when using single precision and/or 512-bit AVX registers. In the design
of our kernel, we will try to reuse values of $A$ instead of $C$ and $S$.

Our kernel will apply $n$ waves of $k_r$ rotations to $m_r$ rows of $A$. One
wave touches $k_r + 1$ columns of $A$, of which $k_r$ can be reused for the
next wave of rotations. The number of memory operations needed for one block is
now
\begin{equation}\label{eq: kernelmemops}
    (\frac{2}{k_r} + \frac{2}{n_b} + \frac{2}{m_r} )m_b(n_b-k_b)k_b \text{  memory operations.}
\end{equation}
This equation is similar to Equation~\eqref{eq:memops_fused} so it may seem
that we have not improved the number of memory operations, but the key is that
$m_r$ can be much larger than $2$. Assuming we have 16 256-bit AVX registers
and are working in double precision, we can choose $m_r = 8$ and $k_r = 5$. If
$n_b$ is sufficiently large, this will reduce the number of memory operations
to
\begin{equation}
    0.65 m(n-k)k \text{  memory operations.}
\end{equation}
This is a factor 3 improvement over the $2 m(n-k)k$ memory operations needed by
the $2 \times 2$ fused rotations. We will later show in the experiments that a
kernel with $m_r = 16$, $k_r = 2$ performs slightly better, despite requiring
more memory operations than the $m_r = 8$, $k_r = 5$ kernel.

We could also have chosen to construct a kernel that applies a sequence of
$n_r$ rotations to $m_r$ rows of $A$. This would have resulted in a similar
number of memory operations and similar kernel sizes.

\section{Packing}\label{section:packing}

In the previous sections, we split the algorithm into smaller blocks to allow
for better cache locality and we also rearranged the inner loop to allow for
more register reuse. However, both of these optimizations have the effect of
making the memory accesses less contiguous. Among other reasons, accessing
memory in a contiguous way is important because of cache lines and the
translation lookaside buffer (TLB).

\subsection{Cache lines} When we access an element from the main memory, that
element is put into the cache in case we need it again later. However, the
cache does not track individual elements, but rather cache lines. These are
blocks of typically 64 contiguous bytes. If we access an element, the entire
cache line is loaded into the cache. If we then access another element in the
same cache line, we can get it directly from the cache, which is much faster
than getting it from the main memory. When incrementing $i$ first, we access
the elements of $A$ in a contiguous way, which almost guarantees that each
cache line will be fully utilized. When incrementing $j$ or $p$ first, we only
access $m_r$ values of $A$ in a contiguous way before we move on to the next
column of $A$, which is almost certainly in a different cache line\footnote{If
the matrix is properly aligned to a cache line boundary, its leading dimension
is a multiple of the cache line size and $m_r$ is also a multiple of the cache
line size, then the cache lines will still be fully utilized.}.

\subsection{TLB} An important component of computer architectures that we have
not yet taken into account is virtual memory. If you allocate space for a large
matrix, you can write your program as if the matrix is stored in a contiguous
block of memory, but in reality, the memory is fragmented into chunks called
pages. Pages are typically 4kb. Programmers typically do not need to worry
about translating the virtual (contiguous) memory addresses to the physical
(fragmented) memory addresses because the operating system takes care of this
for you. However, it is important to remember that this translation involves
looking up the physical address in the page table. As mentioned before, memory
lookups are slow, so just as recently used entries of the matrix are stored in
a cache, recently used entries of the page table are stored in the TLB. If the
matrix is large enough so that subsequent columns of $A$ are in different pages
and there are not enough TLB entries to track all the columns of $A$, then
every $m_r$ accesses of $A$ will incur a TLB miss.

% \subsection{NUMA} Lastly, we have to consider NUMA. NUMA stands for\\ non-uniform memory access. In a NUMA system, the memory is divided into multiple nodes, and each core has a different latency to each node. If the matrix is divided among the nodes, then accessing the matrix in a contiguous way will result in fewer memory accesses to other nodes. This is important because accessing memory on another node is much slower than accessing memory on the same node.

\subsection{Packing}

In their discussion of high-performance matrix-matrix multiplications,
Kazushige Goto et al.~\cite{goto2008anatomy} discussed these issues and
introduced the concept of packing. The idea is to make a ``packed'' copy of the
matrix. Instead of being stored in column- or row-major order, this packed copy
is stored in the exact way that it will be accessed. This way, they can fully
utilize the cache lines and avoid TLB misses.

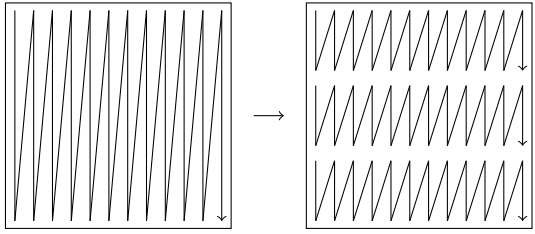
\begin{figure}
    \centering
    \begin{tikzpicture}
        \draw (0,0) rectangle (3,3);
        \foreach \x in {0,...,10}
        \draw(\x/4 + 0.5/4,2.9) -- (\x/4 + 0.5/4,0.1);

        \draw[->] (11/4 + 0.5/4,2.9) -- (11/4 + 0.5/4,0.1);

        \foreach \x in {0,...,10}
        \draw(\x/4 + 0.125,0.1) -- (\x/4 + 0.375,2.9);

        \draw[->] (3.3,1.5) -- (3.7,1.5);

        \begin{scope}[shift={(4,0)}]
            \draw (0,0) rectangle (3,3);

            \foreach \y in {0,...,2}{
            \foreach \x in {0,...,10}
            \draw (\x/4 + 0.5/4,\y+0.9) -- (\x/4 + 0.5/4,\y+0.1);

            \draw[->] (11/4 + 0.5/4,\y+0.9) -- (11/4 + 0.5/4,\y+0.1);
            
            \foreach \x in {0,...,10}
            \draw (\x/4 + 0.125,\y+0.1) -- (\x/4 + 0.375,\y+0.9);
            }

        \end{scope}

    \end{tikzpicture}
    \caption{Illustration of packing for the matrix $A$. The matrix on the left is stored in column-major order, the matrix on the right is stored in packed order.}
    \label{fig:packingA}
\end{figure}

We can apply the same trick to our algorithm. We make a packed copy of $A$,
apply the rotations to the packed copy, and then copy the result back to $A$.
Figure~\ref{fig:packingA} illustrates the packed format. We could also pack the
matrices $C$ and $S$, but this is less important. If we are applying a large
number of rotations, the cost of packing and unpacking the matrix is
negligible. However, if we are applying only a few rotations, the cost of
packing and unpacking the matrix can be significant. If the algorithm is to be
applied to the same matrix multiple times, it may be necessary to keep the
matrix $A$ in packed format instead of repacking on each call.

An extra advantage of packing is that we can make sure the packed matrix is
aligned to a cache line boundary even if the original matrix is not. This is
not only important for cache lines but also for SIMD instructions. Similarly,
if we want to apply rotations from the left instead of from the right (or
equivalently, apply rotations from the right to a row-major matrix), we can
make sure that the accesses are contiguous in the packed matrix.

An important consideration is when to pack and unpack the matrix. To truly have
sequential memory accesses, we should pack and unpack each block. However, the
blocks have a lot of overlap, so we would be packing and unpacking the same
values multiple times. Instead, we pack an entire $m_b \times n$ row-block of
$A$. This leads to some strided accesses but avoids repacking.

\section{Loop orders and choosing block sizes}\label{section:looporder}

We have discussed kernel design and blocking, but we have not discussed the
order of the loops around the kernel and blocks. In this section, we will
decide on the loop order and derive some formulas to choose good block sizes
for the blocked algorithm. Instead of minimizing the number of cache misses at
all levels of the memory hierarchy, we will instead maximize the number of
values that are reused at each cache level. Specifically, we will choose $n_b$
to optimize the L1 cache utilization, $k_b$ to optimize the L2 cache
utilization, and $m_b$ to optimize the L3 cache utilization. We will
specifically optimize for the $m_r = 16$, $k_r = 2$ kernel.  \subsection{The
kernel}

At the lowest level of the algorithm, we have the kernel. This kernel applies
$n_b$ waves of $k_r$ rotations to $m_r$ rows of the matrix. This involves a
block of size $m_r (n_b + k_r)$ for $A$ and of size $n_b k_r$ for $C$ and $S$.
Depending on the loop order, either the values of $A$ or the values of $C$ and
$S$ will be reused, so ideally, these blocks should fit into the L1 cache. We
require all three blocks to fit into the L1 cache even though not all values
will be reused to avoid evicting the wrong values from the cache. For maximum
reuse, we should choose $n_b$ as large as possible.

If we assume the L1 cache can fit $T_1$ floats, then we need to solve the
following equation for $n_b$:
\begin{equation}
    m_r (n_b + k_r) + 2 n_b k_r \leq T_1.
\end{equation}
\begin{equation}
    \Rightarrow  n_b \leq  \frac{T_1 - m_r k_r}{m_r + 2k_r}.
\end{equation}
On our machine, $T_1 = 4000$, so $n_b \leq 220$. To leave some room for other
values, we will choose $n_b = 216$.

Note, if we had designed our kernel differently so that it applies sequences of
$n_r$ rotations instead of a wave of $k_r$ rotations, we would have optimized
$k_b$ instead of $n_b$ for the L1 cache and ended up with a similar equation.
However, it is common to apply the full algorithm with large $m$ and $n$, but
small $k$. This is for example the case when applying delayed sequences of
rotations in the implicit QR algorithm. If $k$ is smaller than $k_b$, we will
not be able to fully utilize the cache.

\subsection{First loop around the kernel}

The next level of the algorithm is the first loop around the kernel. If we
choose to apply the same rotation to other rows of $A$, we can reuse $2 n_b
k_r$ values of $C$ and $S$. If we choose to apply a different set of rotations,
we can reuse $m_r n_b$ values of $A$. Which order leads to more reuse depends
on the size of the kernel. Typically, $m_r$ is going to be larger than $2 k_r$,
so we will choose to reuse values of $A$ over values of $C$ and $S$. This also
makes the packing of $A$ easier.

The loop now applies the kernel $\frac{k_b}{k_r}$ times. Because of the overlap
between the blocks, most of the values of $A$ are reused and remain in the L1
cache. To choose $k_b$, we will make sure that the bigger $m_r (n_b + k_b)$
block of $A$ and $n_b k_b$ blocks of $C$ and $S$ fit into the L2 cache. If we
assume the L2 cache can fit $T_2$ floats, then we need to solve the following
equation for $k_b$:
\begin{equation}
    m_r (n_b + k_b) + 2 n_b k_b \leq T_2.
\end{equation}
\begin{equation}
    k_b \leq \frac{T_2 - m_r n_b}{m_r + 2n_b}.
\end{equation}
Note that this equation depends on $n_b$, which also explains why we chose
$n_b$ first. On our machine, $T_2 = 32000$, so $k_b \leq 62$. We will choose
$k_b = 60$.

\subsection{Second loop around the kernel (full block)}

The next level of the algorithm is the second loop around the kernel, where we
will apply the rotations to the next rows of $A$. The block of $A$ expands from
$m_r (n_b + k_b)$ to $m_b (n_b + k_b)$. We will make sure that this bigger
block fits in the L3 cache. If we assume the L3 cache can fit $T_3$ floats,
then we need to solve the following equation for $m_b$:
\begin{equation}
    m_b (n_b + k_b) \leq T_3.
\end{equation}
\begin{equation}
    m_b \leq \frac{T_3}{n_b + k_b}.
\end{equation}
On our machine, $T_3 = 4480000$, so $m_b \leq 16231$. We will choose a much
smaller value because the L3 cache is shared between all cores and we want to
avoid interfering with other processes. We will choose $m_b = 4800$.

\subsection{Loops around the blocks}

We could similarly select the order of the loops around the blocks to optimize
the memory cost. However, we will choose the order of these loops with an eye
towards practical implementations. For the outer loop, we choose the loop over
$i$ (the row blocks of $A$). This is because it makes the algorithm easier to
parallelize. For the second loop around the blocks, we choose $p$ (the column
blocks of $C$ and $S$) because of the startup and shutdown phases. If we split
$k$ into blocks of $k_b$ first, we need to do $\frac{k}{k_b}$ startup and
shutdown phases of size $k_b \times k_b$, otherwise, we need to do one startup
and shutdown phase of size $k \times k$, which would mean more work is done in
the startup and shutdown phases. In principle, the startup and shutdown phases
can be implemented just as efficiently as the pipeline phase, but it does
require more effort from the programmer. By splitting $k$ into blocks of $k_b$
first, we can get away with a simpler implementation.

\begin{figure}
    \centering
    \subcaptionbox{Third loop around the blocks (full algorithm)}[.99\textwidth]{%
    \begin{tikzpicture}[scale=0.8]
        \draw (0,0) rectangle (5,5);

        \draw[decorate,decoration={brace,amplitude=10pt,mirror,raise=4pt}] (0,5) -- (0,0);
        \node at (-0.9,2.5) {$m$};

        \draw[decorate,decoration={brace,amplitude=10pt,mirror,raise=4pt}] (5,5) -- (0,5);
        \node at (2.5,5.9) {$n$};

        \draw[decorate,decoration={brace,amplitude=10pt,mirror,raise=4pt}] (5,2.5) -- (5,5);
        \node at (5.9,2.5+1.25) {$m_b$};

        \draw (0,2.5) -- (5,2.5);

        \begin{scope}[shift={(8,0)}]
            \draw (0,0) rectangle (1.2,5);

            \draw[decorate,decoration={brace,amplitude=10pt,mirror,raise=4pt}] (0,5) -- (0,0);
            \node at (-1.2,2.5) {$n-1$};

            \draw[decorate,decoration={brace,amplitude=10pt,mirror,raise=4pt}] (1.2,5) -- (0,5);
            \node at (0.6,5.9) {$k$};
            
        \end{scope}
    \end{tikzpicture}
    }
    \subcaptionbox{Second loop around the blocks}[.99\textwidth]{%
    \begin{tikzpicture}[scale=0.8]
        \draw (0,0) rectangle (5,2.5);

        \draw[decorate,decoration={brace,amplitude=10pt,mirror,raise=4pt}] (0,2.5) -- (0,0);
        \node at (-0.9,1.25) {$m_b$};

        \draw[decorate,decoration={brace,amplitude=10pt,mirror,raise=4pt}] (5,2.5) -- (0,2.5);
        \node at (2.5,2.5+0.8) {$n$};

        \begin{scope}[shift={(8,-1)}]
            \draw (0,0) rectangle (1.2,5);

            \draw (0.6,0) -- (0.6,5);

            \draw[decorate,decoration={brace,amplitude=10pt,mirror,raise=4pt}] (0,5) -- (0,0);
            \node at (-1.2,2.5) {$n-1$};

            \draw[decorate,decoration={brace,amplitude=10pt,mirror,raise=4pt}] (1.2,5) -- (0,5);
            \node at (0.6,5.9) {$k$};
            
        \end{scope}
    \end{tikzpicture}
    }
    \subcaptionbox{First loop around the blocks}[.99\textwidth]{%
    \begin{tikzpicture}[scale=0.8]
        \draw (0,0) rectangle (5,2.5);

        \draw[decorate,decoration={brace,amplitude=10pt,mirror,raise=4pt}] (0,2.5) -- (0,0);
        \node at (-0.9,1.25) {$m_b$};

        \draw[decorate,decoration={brace,amplitude=10pt,mirror,raise=4pt}] (5,2.5) -- (0,2.5);
        \node at (2.5,2.5+0.8) {$n$};

        \draw (0.6,0) -- (0.6,2.5);
        \draw (2.12,0) -- (2.12,2.5);
        \draw (2.76,0) -- (2.76,2.5);
        \draw (2.76+2.12,0) -- (2.76+2.12,2.5);

        \draw[pattern=north west lines, pattern color=winered] (0,0) rectangle (2.76,2.5);

        \begin{scope}[shift={(8,-1)}]
            \draw (0,0) rectangle (0.6,5);

            \draw (0,4.4) -- (0.6,5);
            \draw (0,4.4-2.16) -- (0.6,5-2.16);
            \draw (0,4.4-2*2.16) -- (0.6,5-2*2.16);
            \draw (0,0) -- (0.6,0.6);

            \draw[pattern=north west lines, pattern color=winered] (0,4.4) -- (0.6,5) -- (0.6,5-2.16) -- (0,4.4-2.16) -- (0,4.4);

            \draw[decorate,decoration={brace,amplitude=10pt,mirror,raise=4pt}] (0,5) -- (0,0);
            \node at (-1.2,2.5) {$n-1$};

            \draw[decorate,decoration={brace,amplitude=10pt,mirror,raise=4pt}] (0.6,5) -- (0,5);
            \node at (0.3,5.9) {$k_b$};

            \draw[decorate,decoration={brace,amplitude=10pt,mirror,raise=4pt}] (0.6,4.4) -- (0.6,5);
            \node at (1.5,4.7) {$k_b$};

            \draw[decorate,decoration={brace,amplitude=10pt,mirror,raise=4pt}] (0.6,5-2*2.16) -- (0.6,5-2.16);
            \node at (1.5,5-2.16-2.16/2) {$n_b$};
            
        \end{scope}
    \end{tikzpicture}
    }
    \caption{Illustration of the blocking scheme. On the left, the matrix $A$ to whose columns the rotations are applied, and on the right, the matrix $C$ containing the cosines of the rotations. We do not show the matrix $S$ here because its blocks are identical to those of $C$. One of the blocks is indicated with diagonal lines. Notice how that block covers two of the rectangles in $A$: the blocks in $A$ overlap.}
    \label{fig:fullalg}
\end{figure}

\begin{figure}
    \centering
	\subcaptionbox{Second loop around the kernel (full block)}[.99\textwidth]{%
    \begin{tikzpicture}
        \draw[pattern=north west lines, pattern color=winered] (0,0) rectangle (2.76,2.5);

        % show the blocks
        \foreach \x in {0,...,4}
        \draw (0,0.5*\x) rectangle (2.76,0.5*\x+0.5);

        \draw[decorate,decoration={brace,amplitude=10pt,mirror,raise=4pt}] (0,2.5) -- (0,0);
        \node at (-0.8,1.25) {$m_b$};

        \draw[decorate,decoration={brace,amplitude=10pt,mirror,raise=4pt}] (2.76,2) -- (2.76,2.5);
        \node at (3.5,2.25) {$m_r$};

        \draw[decorate,decoration={brace,amplitude=10pt,mirror,raise=4pt}] (2.76,2.5) -- (0,2.5);
        \node at (1.38,2.5+0.8) {$n_b + k_b$};

        \begin{scope}[shift={(5,0)}]

            \draw[pattern=crosshatch,pattern color=orientblue] (0,2.16) -- (0.6,2.76)-- (0.6,0.6) -- (0,0) -- (0,2.16);

            \draw[decorate,decoration={brace,amplitude=10pt,mirror,raise=4pt}] (0,2.16) -- (0,0);
            \node at (-0.8,1.08) {$n_b$};
                
            \draw[decorate,decoration={brace,amplitude=10pt,mirror,raise=4pt}] (0.6,2.76) -- (0,2.76);
            \node at (0.3,3.5) {$k_b$};
        \end{scope}
    \end{tikzpicture}
    }
	\subcaptionbox{First loop around the kernel}[.99\textwidth]{%
    \begin{tikzpicture}
        \draw[pattern=horizontal lines, pattern color=ferngreen] (0,0) rectangle (2.76,0.5);

            % show the blocks
            % \foreach \x in {0,...,4}
            % \draw (0.48-0.12*\x,-0.05-0.05*\x) rectangle (2.76-0.12*\x,0.45-0.05*\x);

        \foreach \x in {0,...,4}
        \draw (0.48-0.12*\x,0) rectangle (2.76-0.12*\x,0.5);

        \draw[decorate,decoration={brace,amplitude=10pt,mirror,raise=4pt}] (0,0.5) -- (0,0);
        \node at (-0.8,0.25) {$m_r$};

        \draw[decorate,decoration={brace,amplitude=10pt,mirror,raise=4pt}] (2.76,0.5) -- (0,0.5);
        \node at (1.38,1.25) {$n_b + k_b$};

        \draw[decorate,decoration={brace,amplitude=10pt,mirror,raise=4pt}] (0.48,0) -- (2.76,0);
        \node at (1.62,-0.8) {$n_b + k_r$};

        \begin{scope}[shift={(5,-2)}]

            \draw[pattern=crosshatch,pattern color=orientblue] (0,2.16) -- (0.6,2.76)-- (0.6,0.6) -- (0,0) -- (0,2.16);

            % show the blocks
            \foreach \x in {0,...,4}
            \draw (0.12*\x,2.16+0.12*\x) -- (0.12*\x+0.12,2.28+0.12*\x)-- (0.12*\x+0.12,0.12+0.12*\x) -- (0.12*\x,0.12*\x) -- (0.12*\x,2.16+0.12*\x);

            \draw[decorate,decoration={brace,amplitude=10pt,mirror,raise=4pt}] (0,2.16) -- (0,0);
            \node at (-0.8,1.08) {$n_b$};
                
            \draw[decorate,decoration={brace,amplitude=10pt,mirror,raise=4pt}] (0.6,2.76) -- (0,2.76);
            \node at (0.3,3.5) {$k_b$};
        \end{scope}
    \end{tikzpicture}
    }
	\subcaptionbox{Kernel}[.99\textwidth]{%
    \begin{tikzpicture}
        \draw[pattern=horizontal lines, pattern color=ferngreen] (0.48,0) rectangle (2.76,0.5);

        \draw[decorate,decoration={brace,amplitude=10pt,mirror,raise=4pt}] (0.48,0.5) -- (0.48,0);
        \node at (-0.3,0.25) {$m_r$};

        \draw[decorate,decoration={brace,amplitude=10pt,mirror,raise=4pt}] (2.76,0.5) -- (0.48,0.5);
        \node at (1.62,1.25) {$n_b + k_r$};

        \begin{scope}[shift={(5,-1.5)}]

            \draw (0,2.16) -- (0.12,2.3)-- (0.12,0.12) -- (0,0) -- (0,2.16);

            \draw[decorate,decoration={brace,amplitude=10pt,mirror,raise=4pt}] (0,2.16) -- (0,0);
            \node at (-0.8,1.08) {$n_b$};
                
            \draw[decorate,decoration={brace,amplitude=10pt,mirror,raise=4pt}] (0.12,2.3) -- (0,2.3);
            \node at (0.06,3.1) {$k_r$};
        \end{scope}

        %draw the legend
        \begin{scope}[shift={(7,0)}]

            \draw[pattern=horizontal lines, pattern color=ferngreen] (0,1) rectangle (0.5,1.2);
            \node at (0.9,1.1) {L1};

            \draw[pattern=crosshatch,pattern color=orientblue] (0,0.5) rectangle (0.5,0.7);
            \node at (0.9,0.6) {L2};

            \draw[pattern=north west lines, pattern color=winered] (0,0) rectangle (0.5,0.2);
            \node at (0.9,0.1) {L3};

        \end{scope}
    \end{tikzpicture}
    }
    \caption{Illustration of the application of a block of rotations using the kernel. On the left, the matrix $A$ to whose columns the rotations are applied, and on the right, the matrix $C$ containing the cosines of the rotations. We do not show the matrix $S$ here because its blocks are identical to those of $C$. }
    \label{fig:fullblock}
\end{figure}
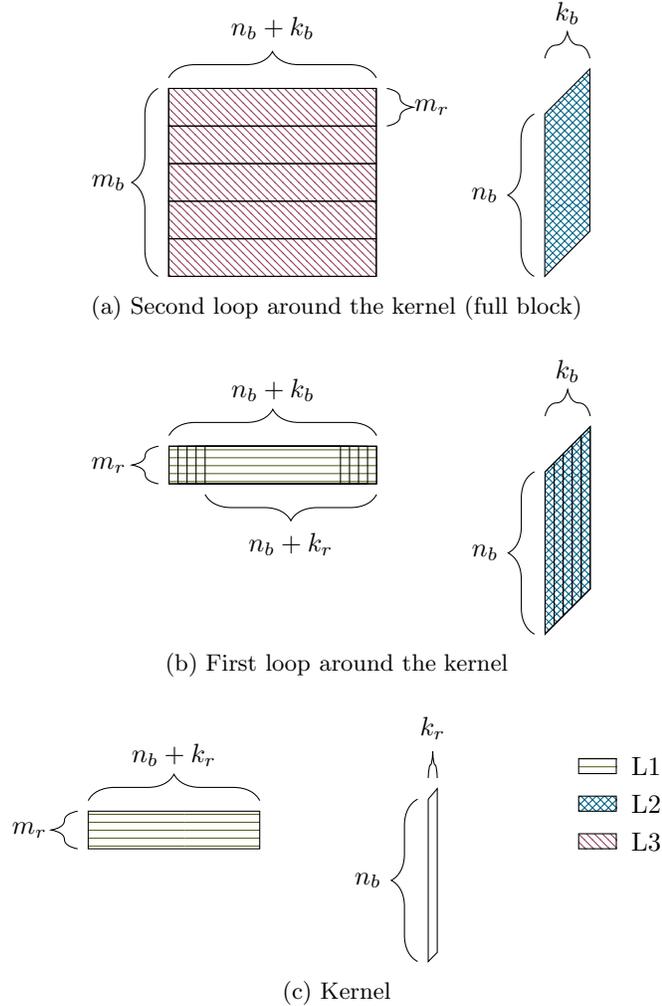

\section{Reducing the number of operations}

In this paper, we have focused on reducing the memory cost of applying a
sequence of Givens rotations. In the literature, there have also been attempts
to reduce the number of floating point operations. Normally, applying a
rotation involves 4 multiplications and 2 additions. By keeping track of a
scaling factor throughout the algorithm, it is possible to reduce this to 2
multiplications and 2 additions. This is called a modified or fast Givens
rotation~\cite{anda1994fast}. We will refer to the paper for the details of the
algorithm. What is important for our purposes is that even though the modified
Givens rotations involve fewer flops, they do require a branch. On modern
architectures, with deep pipelines, branches can be very expensive.

Alternatively, we can use 2x2 reflectors instead of rotations. A 2x2 reflector
can fulfill the same role as a rotation, but it involves 3 multiplications and
3 additions. This is just as many flops as applying a rotation (6 flops), but
it is still advantageous because multiplications are typically more expensive
than additions. Additionally, modern architectures typically include
fused-multiply-add instructions (FMA) that are faster than separate multiply
and add instructions. Naturally, these instructions can only be used when the
number of multiplications is equal to the number of additions. Despite their ability to use FMA instructions effectively, we will show in Section~\ref{sec: experiments} that our implementation of the reflectors is actually slower than the implementation of the rotations. Further research will be needed to determine the cause.

\section{Parallelization}\label{section: parallelization}

There are two places where we can parallelize the algorithm. The first is the
loop over $i_b$, the second is the loop over $i$. The other loops cannot easily
be parallelized.

When parallelizing, we need to take into account which of the caches are
shared. If the L2 cache is shared, we may need to reduce $k_b$ to avoid
interference between the different cores. The L3 cache being shared is not a
problem if we parallelize over $i$, but if we parallelize over $i_b$, we may
need to reduce $m_b$ to avoid interference.

Finally, we also need to make sure the load is balanced between the different
cores. Instead of using a fixed blocksize $m_b$, we can use
$\frac{m}{nthreads}$, rounding up to the nearest multiple of $m_r$. This way,
the load is balanced between the different cores.

\section{Experiments}\label{sec: experiments}

In this final section, we test the performance of the different algorithms on
two different machines. We will compare the following algorithms:
\begin{itemize}
    \item rs\_unoptimized: The naive algorithm to apply a rotation sequence described in Algorithm~\ref{alg:rot_sequence}. 
    \item rs\_blocked: Uses blocks as described in Section~\ref{section:blocking}, but does not use the kernel described in Section~\ref{section:kernel}.
    \item rs\_fused: The algorithm described in~\cite{van2014restructuring}. It uses $2 \times 2$ fused rotations whenever possible (including in the startup and shutdown phases).
    \item rs\_gemm: Instead of applying the rotations directly, this algorithm accumulates blocks of rotations into orthogonal matrices using 2x2 fused rotations and applies those using DGEMM and DTRMM from MKL. Note: this algorithm requires more flops than the other algorithms. When reporting the flop rate, we will only count the flops required to apply the rotations.
    \item rs\_kernel: The algorithm described in this paper. It uses an $m_r = 16$, $k_r = 2$ kernel, but switches to an $m_r = 16$, $k_r = 1$ kernel to apply the startup and shutdown phases\footnote{It is also possible to use a modified $m_r = 16$, $k_r = 2$ kernel for large parts of the startup and shutdown phases, we just did not implement this because of time constraints.}.
    \item rs\_kernel\_v2: Same as rs\_kernel, but the matrix $A$ is already in packed format before the algorithm is called.
\end{itemize}
The algorithms are implemented in C and are (where applicable) parallelized using OpenMP. The experiments are run on two different machines:
\begin{itemize}
    \item Xeon V2: this machine has two Intel Xeon E5-2650 v2 processors, each of which has 8 cores. These processors support AVX instructions but do not have FMA instructions. At the base clock rate, this machine has a peak single-core double precision flop rate of 20.8 Gflop/s. At the maximum clock rate, the flop rate increases to 27.2 Gflop/s.
    \item Xeon V3: this machine has two Intel Xeon E5-2697 v3 processors, each of which has 14 cores. These processors support AVX2 instructions and FMA instructions. At the base clock rate, this machine has a peak single-core double precision flop rate of 41.6 Gflop/s. At the maximum clock rate, the flop rate increases to 54.4 Gflop/s.
\end{itemize}

\subsection{Serial performance}

We first test the performance of the different algorithms on a single core. We
apply the algorithms with $k = 180$, varying $n$ and $m=n$. The results are
shown in Figure~\ref{fig: serial}. The first thing to notice is that the
blocked version and the unoptimized version achieve about the same flop rate
for small matrices, but because of poor cache utilization, the performance of
the unoptimized version quickly drops for even moderately sized matrices. The
blocked version, on the other hand, maintains a high flop rate for all $n$. The
version with 2x2 fusing is approximately 30\% faster than the blocked version
for all $n$. F. Van Zee et al.~\cite{van2014restructuring} reported that their
fused version was 50\% faster than the blocked version on their machine. The
difference between the two speedups is likely because they tested their
algorithm on a different machine. It is also interesting to see that for large
matrices, rs\_gemm outperforms rs\_fused. The poor performance of rs\_gemm for
small matrices shows that the accumulation of the rotations can be an important
bottleneck. This would likely be even more pronounced in a parallelized
version. Finally, we can see that the kernel version is about 60\% faster than
the blocked version and between 20 and 30\% faster than the fused version for
all $n$. We also see that rs\_kernel\_v2 is noticeably faster than rs\_kernel
for large matrices. Even though the difference is small, it is likely worth the
effort to keep the matrix in packed format if possible. Finally, we see that on
the Xeon V2, the flop rate of the kernel version is close to the theoretical
peak flop rate of the machine. On the Xeon V3, we do not reach the peak flop
rate, but we do get closer than with the other algorithms.

\subsection{Selecting kernel size}

\begin{figure}
    \centering

	\subcaptionbox{Xeon V2 flop rate}[.49\textwidth]{%
    \centering
    \begin{tikzpicture}
        \begin{axis}[
                xlabel=$n$,
                ylabel=Gflop/s,
                grid=major,
                xtick={0,1000,2000,3000},
                width=0.49\textwidth,
                ymin=16,
                legend style={at={(0,1.03)},anchor=south west},
                mark size=1.5pt,
                cycle list={
                    {black,mark=none},
                    {ferngreen,mark=square},
                    {perfumepurple,mark=o},
                    {apricotorange,mark=+},
                    {orientblue,mark=triangle},
                    {winered,mark=otimes},
                    {downygreen,mark=*}
                }, 
            ]

            % \addplot table [x=n, y expr={0.75*27.2}, col sep=comma] {anc/testresults_simpson_3/results_12xnx3_prepacked.csv};
            % \addlegendentry{theoretical peak}
            \addplot table [x=n, y=Flops, col sep=comma] {anc/testresults_simpson_3/results_48xnx1_prepacked.csv};
            \addlegendentry{$48\times 1$}
            \addplot table [x=n, y=Flops, col sep=comma] {anc/testresults_simpson_3/results_16xnx2_prepacked.csv};
            \addlegendentry{$16\times 2$}
            \addplot table [x=n, y=Flops, col sep=comma] {anc/testresults_simpson_3/results_12xnx3_prepacked.csv};
            \addlegendentry{$12\times 3$}
            \addplot table [x=n, y=Flops, col sep=comma] {anc/testresults_simpson_3/results_8xnx5_prepacked.csv};
            \addlegendentry{$8\times 5$}

        \end{axis}
    \end{tikzpicture}
    }
    \subcaptionbox{Xeon V3 flop rate}[.49\textwidth]{%
    \centering
    \begin{tikzpicture}
        \begin{axis}[
                xlabel=$n$,
                ylabel=Gflop/s,
                grid=major,
                xtick={0,1000,2000,3000},
                width=0.49\textwidth,
                legend style={at={(0,1.03)},anchor=south west},
                mark size=1.5pt,
                cycle list={
                    {black,mark=none},
                    {ferngreen,mark=square},
                    {perfumepurple,mark=o},
                    {apricotorange,mark=+},
                    {orientblue,mark=triangle},
                    {winered,mark=otimes},
                    {downygreen,mark=*}
                }, 
            ]

            % \addplot table [x=n, y expr={0.75*54.4}, col sep=comma] {anc/testresults_francis_6/results_12xnx3_prepacked.csv};
            \addplot table [x=n, y=Flops, col sep=comma] {anc/testresults_francis_6/results_48xnx1_prepacked.csv};
            \addplot table [x=n, y=Flops, col sep=comma] {anc/testresults_francis_6/results_16xnx2_prepacked.csv};
            \addplot table [x=n, y=Flops, col sep=comma] {anc/testresults_francis_6/results_12xnx3_prepacked.csv};
            \addplot table [x=n, y=Flops, col sep=comma] {anc/testresults_francis_6/results_8xnx5_prepacked.csv};

        \end{axis}
    \end{tikzpicture}
    }
    \caption{Performance of rs\_kernel\_v2 for different block sizes.}
    \label{fig: kernelselection}
\end{figure}
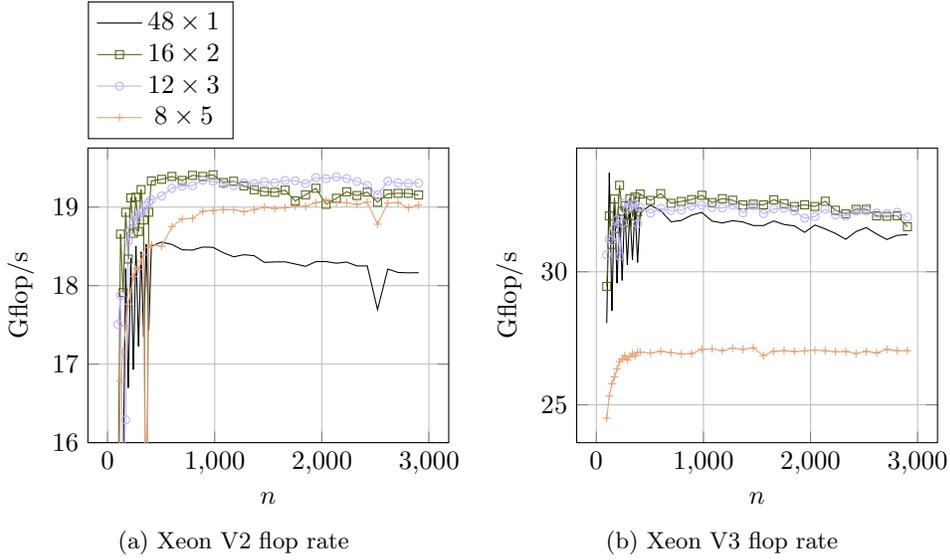

In rs\_kernel, we have chosen the kernel of size $m_r = 16$, $k_r = 2$. We will
now test the performance of the algorithm for different kernel sizes, which
will show that this kernel is the fastest. Since the blocksizes $m_b$, $k_b$,
and $n_b$ are also dependent on the size of the kernel, we have tuned different
blocksizes for each kernel size and use those in the experiments. The results
are shown in Figure~\ref{fig: kernelselection}. We see that the $m_r = 16$,
$k_r = 2$ is indeed the fastest, although the difference with the $m_r = 12$,
$k_r = 3$ kernel is small. A small note here is that we did not fully optimize
the startup and shutdown phases. In those phases, we revert to a kernel with
$k_r = 1$, which means that there is a small bias toward kernels with small
$k_r$. It may well be that the $m_r = 12$, $k_r = 3$ kernel performs better if
the startup and shutdown phases are fully optimized. It is also noteworthy that
according to Equation~\eqref{eq: kernelmemops}, the $m_r = 16$, $k_r = 2$
kernel needs almost twice as many memory operations as the $m_r = 8$, $k_r = 5$
kernel. We do not currently have a satisfying explanation as to why it is still
faster.

\begin{figure}
    \centering

	\subcaptionbox{Xeon V2 flop rate}[.49\textwidth]{%
    \centering
    \begin{tikzpicture}
        \begin{axis}[
                xlabel=$n$,
                ylabel=Gflop/s,
                grid=major,
                xtick={0,1000,2000,3000},
                width=0.49\textwidth,
                legend style={at={(0,1.03)},anchor=south west},
                mark size=1.5pt,
                cycle list={
                    {black,mark=none},
                    {ferngreen,mark=square},
                    {perfumepurple,mark=o},
                    {apricotorange,mark=+},
                    {orientblue,mark=triangle},
                    {winered,mark=otimes},
                    {downygreen,mark=*}
                }, 
            ]

            \addplot table [x=n, y expr={0.75*27.2}, col sep=comma] {anc/testresults_simpson_3/results_16xnx2_prepacked.csv};
            \addlegendentry{75\% of peak}
            \addplot table [x=n, y=Flops, col sep=comma] {anc/testresults_simpson_3/results_16xnx2_prepacked.csv};
            \addlegendentry{rs\_kernel\_v2}
            \addplot table [x=n, y=Flops, col sep=comma] {anc/testresults_simpson_3/results_16xnx2_packed.csv};
            \addlegendentry{rs\_kernel}
            \addplot table [x=n, y=Flops, col sep=comma] {anc/testresults_simpson_3/results_fused.csv};
            \addlegendentry{rs\_fused}
            \addplot table [x=n, y=Flops, col sep=comma] {anc/testresults_simpson_3/results_blocked.csv};
            \addlegendentry{rs\_blocked}
            \addplot table [x=n, y=Flops, col sep=comma] {anc/testresults_simpson_3/results_unoptimized.csv};
            \addlegendentry{rs\_unoptimized}
            \addplot table [x=n, y=Flops, col sep=comma] {anc/testresults_simpson_3/results_gemm.csv};
            \addlegendentry{rs\_gemm}

        \end{axis}
    \end{tikzpicture}
    }
    \subcaptionbox{Xeon V3 flop rate}[.49\textwidth]{%
    \centering
    \begin{tikzpicture}
        \begin{axis}[
                xlabel=$n$,
                ylabel=Gflop/s,
                grid=major,
                xtick={0,1000,2000,3000},
                width=0.49\textwidth,
                legend style={at={(0,1.03)},anchor=south west},
                mark size=1.5pt,
                cycle list={
                    {black,mark=none},
                    {ferngreen,mark=square},
                    {perfumepurple,mark=o},
                    {apricotorange,mark=+},
                    {orientblue,mark=triangle},
                    {winered,mark=otimes},
                    {downygreen,mark=*}
                }, 
            ]

            \addplot table [x=n, y expr={0.75*54.4}, col sep=comma] {anc/testresults_simpson_3/results_16xnx2_prepacked.csv};
            \addplot table [x=n, y=Flops, col sep=comma] {anc/testresults_francis_6/results_16xnx2_prepacked.csv};
            \addplot table [x=n, y=Flops, col sep=comma] {anc/testresults_francis_6/results_16xnx2_packed.csv};
            \addplot table [x=n, y=Flops, col sep=comma] {anc/testresults_francis_6/results_fused.csv};
            \addplot table [x=n, y=Flops, col sep=comma] {anc/testresults_francis_6/results_blocked.csv};
            \addplot table [x=n, y=Flops, col sep=comma] {anc/testresults_francis_6/results_unoptimized.csv};
            \addplot table [x=n, y=Flops, col sep=comma] {anc/testresults_francis_6/results_gemm.csv};

        \end{axis}
    \end{tikzpicture}
    }
	\subcaptionbox{Xeon V2 speedup}[.49\textwidth]{%
    \centering
    \begin{tikzpicture}
        \begin{axis}[
                xlabel=$n$,
                ylabel=relative runtime,
                grid=major,
                xtick={0,1000,2000,3000},
                width=0.49\textwidth,
                ymin=0.9,
                ymax=2, 
                legend style={at={(0,1.03)},anchor=south west},
                mark size=1.5pt,
                cycle list={
                    {black,mark=none},
                    {ferngreen,mark=square},
                    {perfumepurple,mark=o},
                    {apricotorange,mark=+},
                    {orientblue,mark=triangle},
                    {winered,mark=otimes},
                    {downygreen,mark=*}
                }, 
            ]

            \pgfplotstableread[col sep=comma]{anc/testresults_simpson_3/results_16xnx2_prepacked.csv}\firsttable

            \addplot table [x=n, y expr={\thisrow{Flops} / (0.75*27.2)}] \firsttable;

            \pgfplotstableread[col sep=comma]{anc/testresults_simpson_3/results_16xnx2_prepacked.csv}\thistable
            \pgfplotstablecreatecol[create col/copy column from table=\firsttable{Flops}]{Flopsref}\thistable
            \addplot table [x=n, y expr={\thisrow{Flopsref} / \thisrow{Flops}}] \thistable;

            \pgfplotstableread[col sep=comma]{anc/testresults_simpson_3/results_16xnx2_packed.csv}\thistable
            \pgfplotstablecreatecol[create col/copy column from table=\firsttable{Flops}]{Flopsref}\thistable
            \addplot table [x=n, y expr={\thisrow{Flopsref} / \thisrow{Flops}}] \thistable;

            \pgfplotstableread[col sep=comma]{anc/testresults_simpson_3/results_fused.csv}\thistable
            \pgfplotstablecreatecol[create col/copy column from table=\firsttable{Flops}]{Flopsref}\thistable
            \addplot table [x=n, y expr={\thisrow{Flopsref} / \thisrow{Flops}}] \thistable;

            \pgfplotstableread[col sep=comma]{anc/testresults_simpson_3/results_blocked.csv}\thistable
            \pgfplotstablecreatecol[create col/copy column from table=\firsttable{Flops}]{Flopsref}\thistable
            \addplot table [x=n, y expr={\thisrow{Flopsref} / \thisrow{Flops}}] \thistable;

            % \pgfplotstableread[col sep=comma]{anc/testresults_simpson_3/results_unoptimized.csv}\thistable
            % \pgfplotstablecreatecol[create col/copy column from table=\firsttable{Flops}]{Flopsref}\thistable
            % \addplot table [x=n, y expr={\thisrow{Flopsref} / \thisrow{Flops}}] \thistable;

            % \pgfplotstableread[col sep=comma]{anc/testresults_simpson_3/results_gemm.csv}\thistable
            % \pgfplotstablecreatecol[create col/copy column from table=\firsttable{Flops}]{Flopsref}\thistable
            % \addplot table [x=n, y expr={\thisrow{Flopsref} / \thisrow{Flops}}] \thistable;

        \end{axis}
    \end{tikzpicture}
    }
	\subcaptionbox{Xeon V3 speedup}[.49\textwidth]{%
    \centering
    \begin{tikzpicture}
        \begin{axis}[
                xlabel=$n$,
                ylabel=relative runtime,
                grid=major,
                xtick={0,1000,2000,3000},
                width=0.49\textwidth,
                ymin=0.9,
                ymax=2, 
                legend style={at={(0,1.03)},anchor=south west},
                mark size=1.5pt,
                cycle list={
                    {black,mark=none},
                    {ferngreen,mark=square},
                    {perfumepurple,mark=o},
                    {apricotorange,mark=+},
                    {orientblue,mark=triangle},
                    {winered,mark=otimes},
                    {downygreen,mark=*}
                }, 
            ]

            \pgfplotstableread[col sep=comma]{anc/testresults_francis_6/results_16xnx2_prepacked.csv}\firsttable

            \addplot table [x=n, y expr={\thisrow{Flops} / (0.75*54.4)}] \firsttable;

            \pgfplotstableread[col sep=comma]{anc/testresults_francis_6/results_16xnx2_prepacked.csv}\thistable
            \pgfplotstablecreatecol[create col/copy column from table=\firsttable{Flops}]{Flopsref}\thistable
            \addplot table [x=n, y expr={\thisrow{Flopsref} / \thisrow{Flops}}] \thistable;

            \pgfplotstableread[col sep=comma]{anc/testresults_francis_6/results_16xnx2_packed.csv}\thistable
            \pgfplotstablecreatecol[create col/copy column from table=\firsttable{Flops}]{Flopsref}\thistable
            \addplot table [x=n, y expr={\thisrow{Flopsref} / \thisrow{Flops}}] \thistable;

            \pgfplotstableread[col sep=comma]{anc/testresults_francis_6/results_fused.csv}\thistable
            \pgfplotstablecreatecol[create col/copy column from table=\firsttable{Flops}]{Flopsref}\thistable
            \addplot table [x=n, y expr={\thisrow{Flopsref} / \thisrow{Flops}}] \thistable;

            \pgfplotstableread[col sep=comma]{anc/testresults_francis_6/results_blocked.csv}\thistable
            \pgfplotstablecreatecol[create col/copy column from table=\firsttable{Flops}]{Flopsref}\thistable
            \addplot table [x=n, y expr={\thisrow{Flopsref} / \thisrow{Flops}}] \thistable;

            % \pgfplotstableread[col sep=comma]{anc/testresults_francis_6/results_unoptimized.csv}\thistable
            % \pgfplotstablecreatecol[create col/copy column from table=\firsttable{Flops}]{Flopsref}\thistable
            % \addplot table [x=n, y expr={\thisrow{Flopsref} / \thisrow{Flops}}] \thistable;

            % \pgfplotstableread[col sep=comma]{anc/testresults_francis_6/results_gemm.csv}\thistable
            % \pgfplotstablecreatecol[create col/copy column from table=\firsttable{Flops}]{Flopsref}\thistable
            % \addplot table [x=n, y expr={\thisrow{Flopsref} / \thisrow{Flops}}] \thistable;

        \end{axis}
    \end{tikzpicture}
    }
    \caption{On top, the Flop rates of the different algorithms. On the bottom, the runtime of the different algorithms relative to rs\_kernel\_v2.}
    \label{fig: serial}
\end{figure}
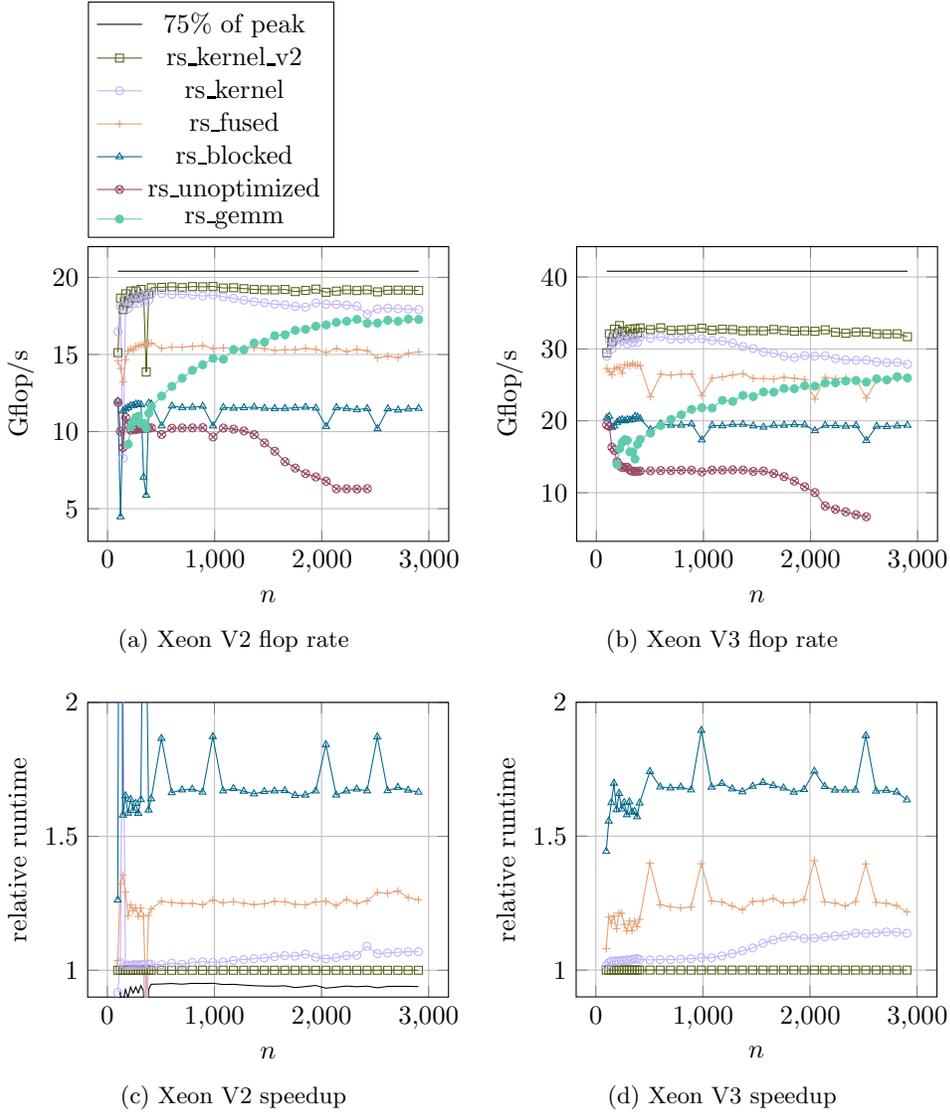

\subsection{Parallel performance}

\begin{figure}
    \centering

	\subcaptionbox{Xeon V2}[.49\textwidth]{%
    \centering
    \begin{tikzpicture}
        \begin{semilogyaxis}[
                xlabel=$n$,
                ylabel=Gflop/s,
                ytick={1,3,10,30,100,300},
                yticklabels={1,3,10,30,100,300},
                grid=major,
                width=0.47\textwidth,
                legend style={at={(0,1.03)},anchor=south west},
                mark size=1.5pt,
                cycle list={
                    {black,mark=none},
                    {ferngreen,mark=square},
                    {perfumepurple,mark=o},
                    {apricotorange,mark=+},
                    {orientblue,mark=triangle},
                    {winered,mark=otimes},
                    {downygreen,mark=*}
                }, 
            ]

            \addplot table [x=n, y=Flops, col sep=comma] {anc/testresults_simpson_3/results_16xnx2_prepacked.csv};
            \addlegendentry{serial}
            \addplot table [x=n, y=Flops, col sep=comma] {anc/testresults_simpson_3/results_16xnx2_prepacked_para_2.csv};
            \addlegendentry{2 threads}
            \addplot table [x=n, y=Flops, col sep=comma] {anc/testresults_simpson_3/results_16xnx2_prepacked_para_4.csv};
            \addlegendentry{4 threads}
            \addplot table [x=n, y=Flops, col sep=comma] {anc/testresults_simpson_3/results_16xnx2_prepacked_para_8.csv};
            \addlegendentry{8 threads}
            \addplot table [x=n, y=Flops, col sep=comma] {anc/testresults_simpson_3/results_16xnx2_prepacked_para_16.csv};
            \addlegendentry{16 threads}

        \end{semilogyaxis}
    \end{tikzpicture}
    }
    \subcaptionbox{Xeon V3}[.49\textwidth]{%
    \centering
    \begin{tikzpicture}
        \begin{semilogyaxis}[
                xlabel=$n$,
                ylabel=Gflop/s,
                ytick={1,3,10,30,100,300,600},
                yticklabels={1,3,10,30,100,300,600},
                grid=major,
                xtick={0,1000,2000,3000},
                width=0.47\textwidth,
                legend style={at={(0,1.03)},anchor=south west},
                mark size=1.5pt,
                cycle list={
                    {black,mark=none},
                    {ferngreen,mark=square},
                    {perfumepurple,mark=o},
                    {apricotorange,mark=+},
                    {orientblue,mark=triangle},
                    {winered,mark=otimes},
                    {downygreen,mark=*}
                }, 
            ]

            \addplot table [x=n, y=Flops, col sep=comma] {anc/testresults_francis_6/results_16xnx2_prepacked.csv};
            \addlegendentry{serial}
            \addplot table [x=n, y=Flops, col sep=comma] {anc/testresults_francis_6/results_16xnx2_prepacked_para_2.csv};
            \addlegendentry{2 threads}
            \addplot table [x=n, y=Flops, col sep=comma] {anc/testresults_francis_6/results_16xnx2_prepacked_para_4.csv};
            \addlegendentry{4 threads}
            \addplot table [x=n, y=Flops, col sep=comma] {anc/testresults_francis_6/results_16xnx2_prepacked_para_8.csv};
            \addlegendentry{8 threads}
            \addplot table [x=n, y=Flops, col sep=comma] {anc/testresults_francis_6/results_16xnx2_prepacked_para_14.csv};
            \addlegendentry{14 threads}
            \addplot table [x=n, y=Flops, col sep=comma] {anc/testresults_francis_6/results_16xnx2_prepacked_para_28.csv};
            \addlegendentry{28 threads}

        \end{semilogyaxis}
    \end{tikzpicture}
    }
	\subcaptionbox{Xeon V2}[.49\textwidth]{%
    \centering
    \begin{tikzpicture}
        \begin{semilogyaxis}[
                xlabel=$n$,
                ylabel=Speedup,
                ytick={1,2,4,8,16},
                yticklabels={1,2,4,8,16},
                grid=major,
                width=0.47\textwidth,
                legend style={at={(0,1.03)},anchor=south west},
                mark size=1.5pt,
                cycle list={
                    {black,mark=none},
                    {ferngreen,mark=square},
                    {perfumepurple,mark=o},
                    {apricotorange,mark=+},
                    {orientblue,mark=triangle},
                    {winered,mark=otimes},
                    {downygreen,mark=*}
                }, 
            ]

            \pgfplotstableread[col sep=comma]{anc/testresults_simpson_3/results_16xnx2_prepacked.csv}\firsttable

            \pgfplotstableread[col sep=comma]{anc/testresults_simpson_3/results_16xnx2_prepacked.csv}\thistable
            \pgfplotstablecreatecol[create col/copy column from table=\firsttable{Flops}]{Flopsref}\thistable
            \addplot table [x=n, y expr={\thisrow{Flops} / \thisrow{Flopsref}}] \thistable;

            \pgfplotstableread[col sep=comma]{anc/testresults_simpson_3/results_16xnx2_prepacked_para_2.csv}\thistable
            \pgfplotstablecreatecol[create col/copy column from table=\firsttable{Flops}]{Flopsref}\thistable
            \addplot table [x=n, y expr={\thisrow{Flops} / \thisrow{Flopsref}}] \thistable;

            \pgfplotstableread[col sep=comma]{anc/testresults_simpson_3/results_16xnx2_prepacked_para_4.csv}\thistable
            \pgfplotstablecreatecol[create col/copy column from table=\firsttable{Flops}]{Flopsref}\thistable
            \addplot table [x=n, y expr={\thisrow{Flops} / \thisrow{Flopsref}}] \thistable;

            \pgfplotstableread[col sep=comma]{anc/testresults_simpson_3/results_16xnx2_prepacked_para_8.csv}\thistable
            \pgfplotstablecreatecol[create col/copy column from table=\firsttable{Flops}]{Flopsref}\thistable
            \addplot table [x=n, y expr={\thisrow{Flops} / \thisrow{Flopsref}}] \thistable;

            \pgfplotstableread[col sep=comma]{anc/testresults_simpson_3/results_16xnx2_prepacked_para_16.csv}\thistable
            \pgfplotstablecreatecol[create col/copy column from table=\firsttable{Flops}]{Flopsref}\thistable
            \addplot table [x=n, y expr={\thisrow{Flops} / \thisrow{Flopsref}}] \thistable;

        \end{semilogyaxis}
    \end{tikzpicture}
    }
	\subcaptionbox{Xeon V3}[.49\textwidth]{%
    \centering
    \begin{tikzpicture}
        \begin{semilogyaxis}[
                xlabel=$n$,
                ylabel=Speedup,
                ytick={1,2,4,8,16},
                yticklabels={1,2,4,8,16},
                grid=major,
                width=0.47\textwidth,
                legend style={at={(0,1.03)},anchor=south west},
                mark size=1.5pt,
                cycle list={
                    {black,mark=none},
                    {ferngreen,mark=square},
                    {perfumepurple,mark=o},
                    {apricotorange,mark=+},
                    {orientblue,mark=triangle},
                    {winered,mark=otimes},
                    {downygreen,mark=*}
                }, 
            ]

            \pgfplotstableread[col sep=comma]{anc/testresults_francis_6/results_16xnx2_prepacked.csv}\firsttable

            \pgfplotstableread[col sep=comma]{anc/testresults_francis_6/results_16xnx2_prepacked.csv}\thistable
            \pgfplotstablecreatecol[create col/copy column from table=\firsttable{Flops}]{Flopsref}\thistable
            \addplot table [x=n, y expr={\thisrow{Flops} / \thisrow{Flopsref}}] \thistable;

            \pgfplotstableread[col sep=comma]{anc/testresults_francis_6/results_16xnx2_prepacked_para_2.csv}\thistable
            \pgfplotstablecreatecol[create col/copy column from table=\firsttable{Flops}]{Flopsref}\thistable
            \addplot table [x=n, y expr={\thisrow{Flops} / \thisrow{Flopsref}}] \thistable;

            \pgfplotstableread[col sep=comma]{anc/testresults_francis_6/results_16xnx2_prepacked_para_4.csv}\thistable
            \pgfplotstablecreatecol[create col/copy column from table=\firsttable{Flops}]{Flopsref}\thistable
            \addplot table [x=n, y expr={\thisrow{Flops} / \thisrow{Flopsref}}] \thistable;

            \pgfplotstableread[col sep=comma]{anc/testresults_francis_6/results_16xnx2_prepacked_para_8.csv}\thistable
            \pgfplotstablecreatecol[create col/copy column from table=\firsttable{Flops}]{Flopsref}\thistable
            \addplot table [x=n, y expr={\thisrow{Flops} / \thisrow{Flopsref}}] \thistable;

            \pgfplotstableread[col sep=comma]{anc/testresults_francis_6/results_16xnx2_prepacked_para_14.csv}\thistable
            \pgfplotstablecreatecol[create col/copy column from table=\firsttable{Flops}]{Flopsref}\thistable
            \addplot table [x=n, y expr={\thisrow{Flops} / \thisrow{Flopsref}}] \thistable;

            \pgfplotstableread[col sep=comma]{anc/testresults_francis_6/results_16xnx2_prepacked_para_28.csv}\thistable
            \pgfplotstablecreatecol[create col/copy column from table=\firsttable{Flops}]{Flopsref}\thistable
            \addplot table [x=n, y expr={\thisrow{Flops} / \thisrow{Flopsref}}] \thistable;

        \end{semilogyaxis}
    \end{tikzpicture}
    }
    \caption{On top: the flop rates of rs\_kernel\_v2 using a varying amount of threads. On the bottom: the speedup of the parallel versions relative to the serial version.}
    \label{fig: parallel}
\end{figure}
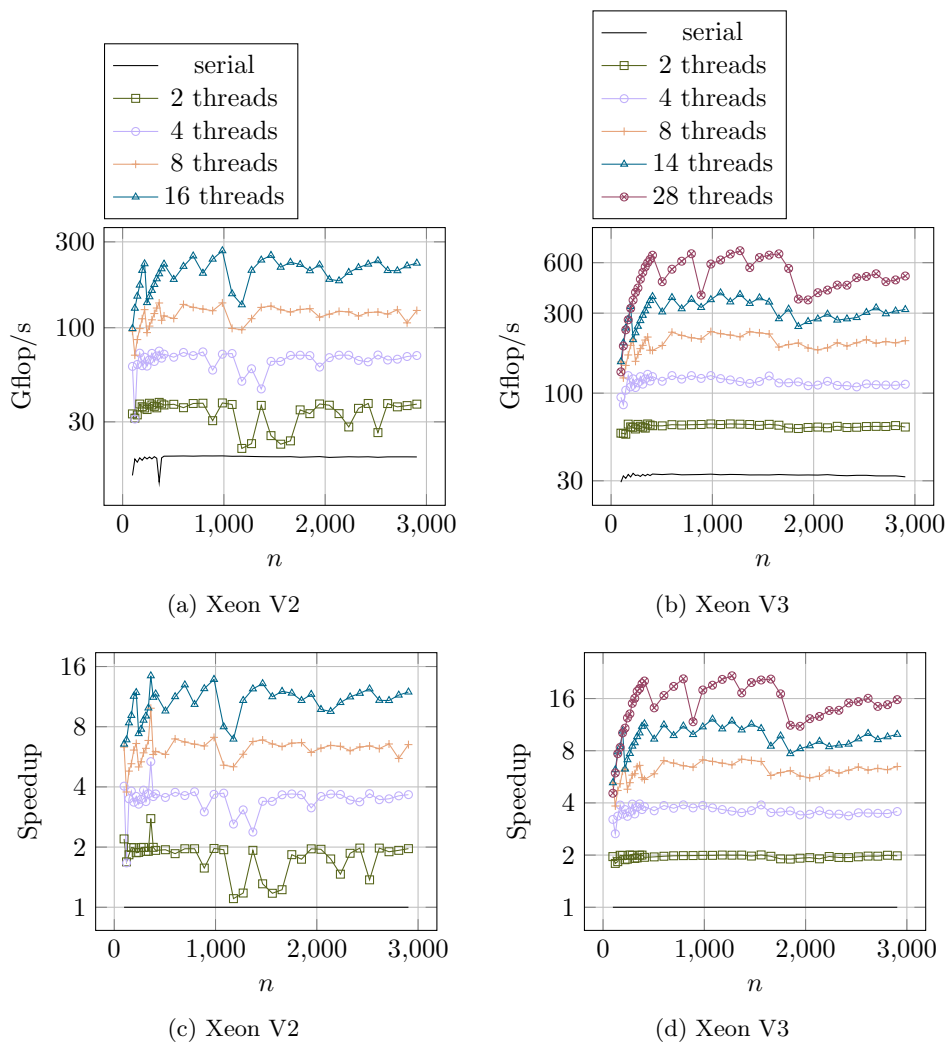

Next, we test the performance of the rs\_kernel\_v2 on multiple cores. We use
the same parameters as before $k = 180$, varying $n$ and $m=n$. We have
parallelized our code around the outer loop ($i_b$). The results are shown in
Figure~\ref{fig: parallel}. The first thing to notice is that the speedup is
quite good. For the Xeon V2, we get a speedup of about 10 for 16 threads. For
the Xeon V3, we get a speedup of about 16 for 28 threads. The high parallel
efficiency is expected because the communication between threads is limited.
The threads apply the same rotations to different rows of $A$. The second thing
to notice is that while the flop rate of the serial code is fairly independent
of $n$, the flop rate of the parallel code goes up and down. This is because of
load balancing. For the kernel to work, we need to have a multiple of $m_r$
rows in each block. That means that if we want each thread to handle the same
number of rows, we need $m$ to be a multiple of $m_r$ times the number of
threads. The peaks in the flop rate are where this is the case. A possible
solution is to make a smaller kernel to handle the edge cases.

\subsection{Performance of 2x2 reflectors}

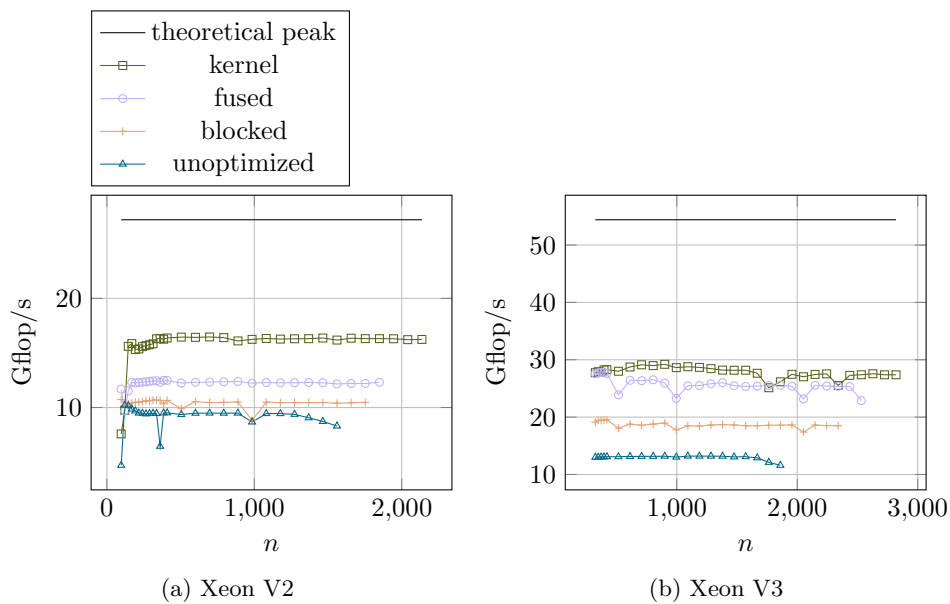
\begin{figure}
    \centering

	\subcaptionbox{Xeon V2}[.49\textwidth]{%
    \centering
    \begin{tikzpicture}
        \begin{axis}[
                xlabel=$n$,
                ylabel=Gflop/s,
                grid=major,
                width=0.49\textwidth,
                legend style={at={(0,1.03)},anchor=south west},
                mark size=1.5pt,
                cycle list={
                    {black,mark=none},
                    {ferngreen,mark=square},
                    {perfumepurple,mark=o},
                    {apricotorange,mark=+},
                    {orientblue,mark=triangle},
                    {winered,mark=otimes},
                    {downygreen,mark=*}
                }, 
            ]

            \addplot table [x=n, y expr={27.2}, col sep=comma] {anc/testresults_simpson_v2/refl/results_12xnx2_packed.csv};
            \addlegendentry{theoretical peak}
            \addplot table [x=n, y=Flops, col sep=comma] {anc/testresults_simpson_v2/refl/results_12xnx2_packed.csv};
            \addlegendentry{kernel}
            \addplot table [x=n, y=Flops, col sep=comma] {anc/testresults_simpson_v2/refl/results_fused.csv};
            \addlegendentry{fused}
            \addplot table [x=n, y=Flops, col sep=comma] {anc/testresults_simpson_v2/refl/results_blocked.csv};
            \addlegendentry{blocked}
            \addplot table [x=n, y=Flops, col sep=comma] {anc/testresults_simpson_v2/refl/results_unoptimized.csv};
            \addlegendentry{unoptimized}

        \end{axis}
    \end{tikzpicture}
    }
    \subcaptionbox{Xeon V3}[.49\textwidth]{%
    \centering
    \begin{tikzpicture}
        \begin{axis}[
                xlabel=$n$,
                ylabel=Gflop/s,
                grid=major,
                width=0.49\textwidth,
                ytick={10,20,30,40,50},
                legend style={at={(0,1.03)},anchor=south west},
                mark size=1.5pt,
                cycle list={
                    {black,mark=none},
                    {ferngreen,mark=square},
                    {perfumepurple,mark=o},
                    {apricotorange,mark=+},
                    {orientblue,mark=triangle},
                    {winered,mark=otimes},
                    {downygreen,mark=*}
                }, 
            ]

            \addplot table [x=n, y expr={54.4}, col sep=comma] {anc/testresults_francis_v2/refl/results_12xnx2_packed.csv};
            \addplot table [x=n, y=Flops, col sep=comma] {anc/testresults_francis_v2/refl/results_12xnx2_packed.csv};
            \addplot table [x=n, y=Flops, col sep=comma] {anc/testresults_francis_v2/refl/results_fused.csv};
            \addplot table [x=n, y=Flops, col sep=comma] {anc/testresults_francis_v2/refl/results_blocked.csv};
            \addplot table [x=n, y=Flops, col sep=comma] {anc/testresults_francis_v2/refl/results_unoptimized.csv};

        \end{axis}
    \end{tikzpicture}
    }
    \caption{Flop rates of different algorithms to apply a sequence of 2x2 reflectors to a matrix. The theoretical peak flop rate is 100\% of the peak flop rate at the maximum frequency.}
    \label{fig: reflectors}
\end{figure}

Finally, we look at the performance of the different algorithms if they are
modified to apply 2x2 reflectors instead of Givens rotations. We use the same
parameters as before: $k = 180$, varying $n$ and $m=n$. The fused algorithm
still uses 2x2 fusing, but the size of the kernel is reduced to $m_r = 12$,
$k_r = 2$. The results are shown in Figure~\ref{fig: reflectors}. Even though
our kernel algorithm is still faster than the other algorithms, it is clear
that switching to 2x2 reflectors negatively impacted the performance.
We emphasize that we do not believe that this indicates a fundamental issue
with using reflectors, but it is rather an indication that further research
will be required to unlock the full potential of the reflectors.

\section{Conclusion}

We have presented a new algorithm to apply sequences of rotations to a matrix.
We have shown that our algorithm is faster than the state-of-the-art and
achieves close to the theoretical limits of the hardware. We have also shown
that our algorithm scales well to multiple cores.

In the conclusion of their paper on the wavefront algorithm, F. G. Van Zee et
al.~\cite{van2014restructuring} noted that their work may result in a
renaissance for the implicit QR algorithm. Although these hopes did not fully
materialize in the 10 years since then, we believe that our work represents
another chance for the implicit QR algorithm. In future work, we will
\begin{itemize}
\item Push for the inclusion of rotation sequences in the BLAS \cite{dongarra1990set}.
Additionally, we will contribute our optimized implementation to BLIS \cite{BLIS1}.
We believe that only through easy access to highly optimized implementations,
other researchers will
be able to take advantage of our work.
\item Investigate the performance of our algorithm on other architectures. It
should be easy to implement an efficient kernel for more recent CPUs with
AVX512 support, but we are also interested in extending the algorithm to GPUs
and distributed systems.
\item Modify existing methods, such as the implicit QR algorithm, to take
advantage of our algorithm. This may present new challenges, such as the need
to handle 3x3 reflectors for the double-shift Hessenberg QR algorithm.
\end{itemize}

\section*{Acknowledgments} Julien was partially supported for this work by NSF
award \#2004850.

\pagebreak

\bibliographystyle{siamplain}
\bibliography{references}

\begin{thebibliography}{10}

\bibitem{anda1994fast}
{\sc A.~A. Anda and H.~Park}, {\em Fast plane rotations with dynamic scaling},
  SIAM Journal on Matrix Analysis and Applications, 15 (1994), pp.~162--174.

\bibitem{dongarra1990set}
{\sc J.~J. Dongarra, J.~Du~Croz, S.~Hammarling, and I.~S. Duff}, {\em A set of
  level 3 basic linear algebra subprograms}, ACM Transactions on Mathematical
  Software (TOMS), 16 (1990), pp.~1--17.

\bibitem{francis1961qr}
{\sc J.~G. Francis}, {\em The {QR} transformation a unitary analogue to the
  {LR} transformation—part 1}, The Computer Journal, 4 (1961), pp.~265--271.

\bibitem{goto2008anatomy}
{\sc K.~Goto and R.~A. v.~d. Geijn}, {\em Anatomy of high-performance matrix
  multiplication}, ACM Transactions on Mathematical Software (TOMS), 34 (2008),
  pp.~1--25.

\bibitem{jacobi1846leichtes}
{\sc C.~G.~J. Jacobi}, {\em {{\"U}ber ein leichtes verfahren die in der theorie
  der s{\"a}cularst{\"o}rungen vorkommenden gleichungen numerisch
  aufzul{\"o}sen}}, Journal für die reine und angewandte Mathematik,  (1846).

\bibitem{kaagstrom2008blocked}
{\sc B.~K{\aa}gstr{\"o}m, D.~Kressner, E.~S. Quintana-Ort{\'\i}, and
  G.~Quintana-Ort{\'\i}}, {\em Blocked algorithms for the reduction to
  {Hessenberg}-triangular form revisited}, BIT Numerical Mathematics, 48
  (2008), pp.~563--584.

\bibitem{olivry2020automated}
{\sc A.~Olivry, J.~Langou, L.-N. Pouchet, P.~Sadayappan, and F.~Rastello}, {\em
  Automated derivation of parametric data movement lower bounds for affine
  programs}, in Proceedings of the 41st ACM SIGPLAN Conference on Programming
  Language Design and Implementation, PLDI 2020, ACM New York, NY, USA, 2020,
  pp.~808–--822.

\bibitem{schreiber1989storage}
{\sc R.~Schreiber and C.~Van~Loan}, {\em A storage-efficient {WY}
  representation for products of {Householder} transformations}, SIAM Journal
  on Scientific and Statistical Computing, 10 (1989), pp.~53--57.

\bibitem{BLIS1}
{\sc F.~G. {V}an {Z}ee and R.~A. {v}an~{d}e {G}eijn}, {\em {BLIS}: A framework
  for rapidly instantiating {BLAS} functionality}, ACM Transactions on
  Mathematical Software, 41 (2015), pp.~14:1--14:33,
  \url{https://doi.acm.org/10.1145/2764454}.

\bibitem{van2014restructuring}
{\sc F.~G. Van~Zee, R.~A. van~de Geijn, and G.~Quintana-Ort{\'\i}}, {\em
  Restructuring the tridiagonal and bidiagonal {QR} algorithms for
  performance}, ACM Transactions on Mathematical Software (TOMS), 40 (2014),
  pp.~1--34.

\end{thebibliography}

\end{document}